%% file: qqbase5.tex
\documentclass[12pt]{article}
\usepackage{psfig, amsfonts}
\usepackage{latexsym}
\input{texdefs.tex}

\input{epsf.tex}
\makeglossary
\title{Growth Rates in the Quaquaversal Tiling\footnote{AMS 1991 Subject
Classification: 52C22, also 20H15, 51F25, 60D05, 60J15, 82D20}
}
\author{Brimstone Draco\footnote{Department of Mathematics, University
of Texas, Austin, TX 78712, brim@math.utexas.edu}
 \and Lorenzo Sadun\footnote{Department of Mathematics, University
of Texas, Austin, TX 78712, sadun@math.utexas.edu}
 \and Douglas Van Wieren\footnote{Department of Computer Science and 
Engineering, Pennsylvania 
State University, University Park, PA 16802, dvw@buchni.bk.psu.edu}}
\date{\today}

\begin{document}
\setsource{qqbase}{Lorenzo}{Version 4}{Almost done!}

\providecommand{\Rx}{\ensuremath{{\bf R}_x}}
\providecommand{\Ry}{\ensuremath{{\bf R}_y}}
\providecommand{\Rz}{\ensuremath{{\bf R}_z}}
\def\call{{\cal L}}
\renewcommand{\Reals}{{\Bbb R}}
\newcommand{\myspecref}[2][\myfilename]{\mbox{(form \ref{\myfilename:Equation:#2})}}


\numberequationsastheorems
\maketitle

\begin{abstract}
   Conway and Radin's ``quaquaversal'' tiling of $\Reals^3$ is known
   to exhibit statistical rotational symmetry in the infinite volume
   limit.  A finite patch, however, cannot be perfectly isotropic, and
   we compute the rates at which the anisotropy scales with size.  In
   a sample of volume $N$, tiles appear in $O(N^{1/6})$ distinct
   orientations.  However, the orientations are not uniformly
   populated. A small ($O(N^{1/84})$) set of these orientations
   account for the majority of the tiles.  Furthermore, these
   orientations are not uniformly distributed on $SO(3)$. Sample
   averages of functions on $SO(3)$ seem to approach their ergodic limits
   as $N^{-1/336}$.  Since even macroscopic patches of a quaquaversal
   tiling maintain noticable anisotropy, a hypothetical physical
   quasicrystal whose structure was similar to the quaquaversal tiling
   could be identified by anisotropic features of its electron
   diffraction pattern.

\end{abstract}

\mysection{Introduction and Results}{intro}

   This paper concerns statistical properties of Conway and Radin's
   \cite{CR} quaquaversal (QQ) tiling of Euclidean 3-space.  A
   quaquaversal tiling is made by copies of a single wedge-shaped
   tile, appearing in an infinite number of distinct orientations.
   The set of orientations is a dense subgroup of $SO(3)$, and the
   orientations are in fact uniformly distributed on $SO(3)$: Given a
   smooth function $f$ on $SO(3)$, the average value of $f$ on the
   orientations of tiles in a large patch of $\Reals^3$ is close to
   the integral of $f$ over $SO(3)$ (with respect to Haar measure).
   This property, called ``statistical rotational symmetry'' by Radin
   \cite{R1}, suggests that a large quasicrystal based on the
   quaquaversal tiling should appear round, and its physical
   properties should be isotropic. We call such a material
   ``asymptotically round''.

\medskip
\centerline{\epsfysize=4truein\epsfbox{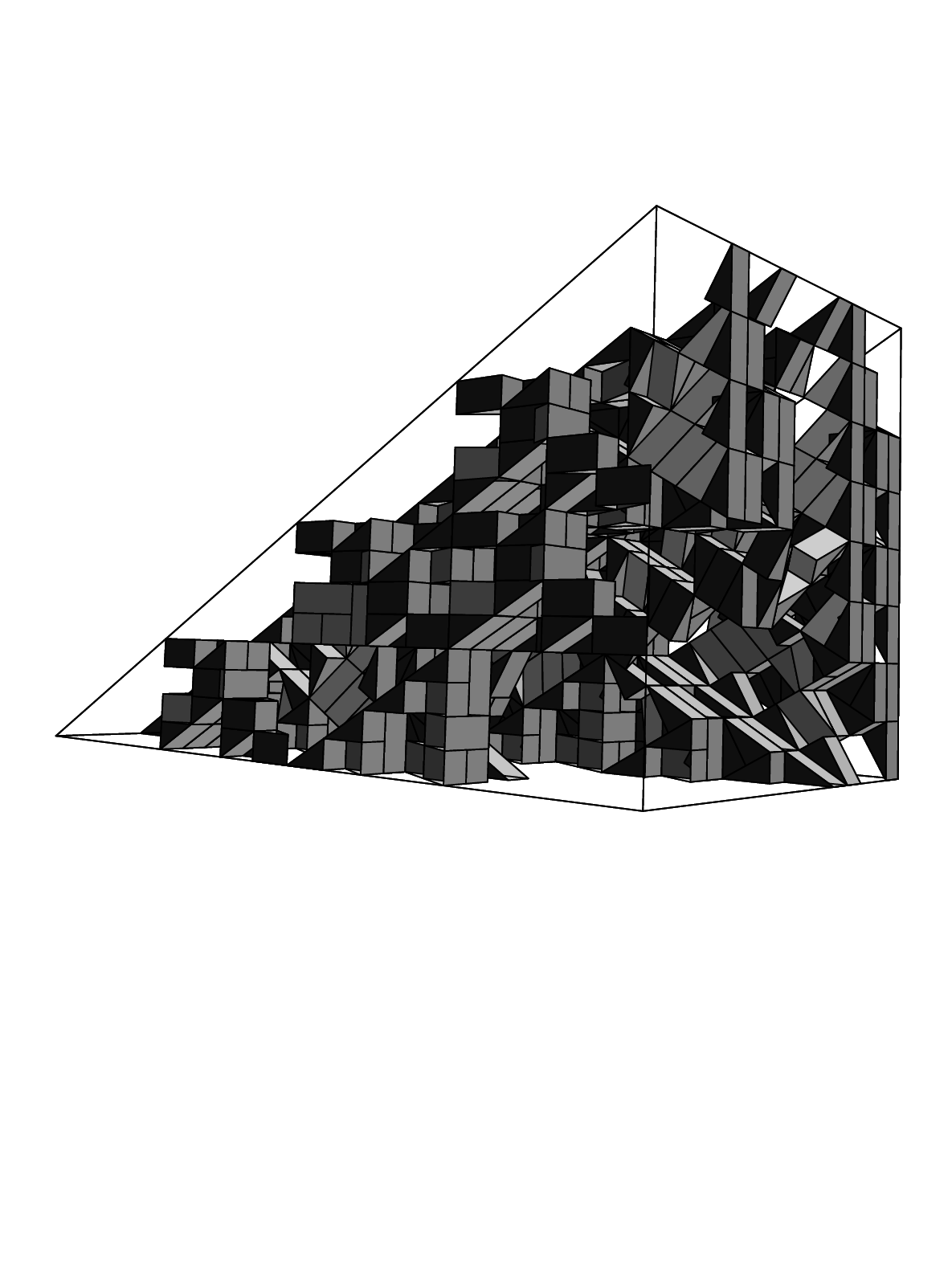}}
\centerline{\bf Figure 1: Part of a QQ tiling}
\smallskip


   In this paper we address the question of how large such a
   quasicrystal would have to be to make its roundness manifest.  We
   show that the convergence to rotational symmetry is polynomial in
   the sample size, but with an extremely small exponent.  Samples
   with $10^{24}$ tiles have pronounced anisotropy, and even samples
   with $10^{100}$ tiles have noticable anisotropy.  In short, the
   asymptotic roundness of the quaquaversal tiling, while
   mathematically correct, is well beyond the range of any conceivable
   physics. If a physical material were modeled on the quaquaversal
   tiling, it would exhibit considerably more isotropy than a crystal,
   while still being considerably less isotropic than a liquid or a
   glass.  These comparisons are discussed further in \myref{Section}{Conc}.
   There is some evidence \cite{R2} that such
   asymptotically round quasicrystals may indeed exist in nature.

   The slow growth of complexity is not unique to the QQ tiling, but
   is a typical feature of 3 dimensional substitution tilings.  Two
   dimensional substitution tilings have even slower growth, with the
   number of orientations growing only as the logarithm of the sample
   size.  We expect that any 3-dimensional substutution tiling that is
   qualitatively more isotropic than the QQ tiling would have to have
   a very complicated substitution rule.  In other words, much of the
   complexity of the distribution of orientations would have to be
   built in from the start, sacrificing the elegant simplicity of the
   quaquaversal tiling.

The orientations of a randomly chosen tile within the QQ tiling is
described by a biased random walk in a dense subgroup of
$SO(3)$. Lubotsky, Phillips and Sarnak \cite{LPS1, LPS2} 
considered {\it un}\/biased random walks
on such subgroups in their study of optimal distributions of points on
$S^2$, and some of our methods are related to theirs. 
Long time behavior of unbiased random walks on infinite
cyclic subgroups of $SO(2)$ have been studied by Su \cite{Su}.

It is possible to compare the optimal distribution of elements of
$SO(3)$ of \cite{LPS1} with the distribution of orientations in the
quaquaversal tiling as follows.  The ``operator discrepancy''
considered in \cite{LPS1} is the operator norm of a certain Hecke
operator acting on a subspace of $L^2(SO(3))$ (or $L^2(S^2)$).  This
operator is Hermitian and has discrete spectrum, so its operator norm
is the supremum of the eigenvalues.  For an optimal distribution of
$N$ points, with $N-1$ prime, this supremum was shown to be
$2\sqrt{N-1}/N$, and in particular to decay as $N^{-1/2}$. In our case
a distribution of orientations of $N$ QQ tiles is obtained by applying
a fixed Hecke operator $\log_8N$ times; the $\log_8N$-th power of the
largest eigenvalue of this Hecke operator goes as $N^{-1/336}$. These
eigenvalues are discussed at greater length in
\myref{Section}{Apperg}.

The operator discrepancy is a very strict measure of anisotropy. Indeed,
all distributions of points on the circle have operator discrepancy 
equal to 1. In this paper we also 
consider two much weaker measures of spreading.
The first is just a count of the number of distinct orientations in $SO(3)$,
which turns out to scale as $N^{1/6}$.
The second is the number of distinct orientations needed to account for
half the sample, which scales as $N^{1/84}$. These measures are discussed
in \myref{Section}{group} and \myref{Section}{PD}, respectively.   

   The QQ tiling is constructed as follows (see \cite{CR} for details).  We
begin with a single right triangular prism, $\sqrt{3}$ units wide, one
unit high, and one unit deep, as in Figure 2a.  This prism is divided
into eight congruent pieces, each similar to the original.  The front
four resulting tiles are shown in Figure 2b, the back four in Figure
2c.  We then rescale the entire pattern by a linear factor of 2,
obtaining a cluster of 8 tiles, each congruent to the original.  

We can repeat this process of subdivision and rescaling to obtain a cluster
of 64 tiles, of 512 tiles, and so on.  We embed each cluster of size $8^n$ 
within the succeeding cluster of size $8^{n+1}$ (there are 8 ways to do this).
The union of all the clusters is then an infinite tiling, typically 
covering all of $\Reals^3$.

   After a ``parent'' tile is subdivided, the orientations of the
   ``daughters'' are closely related to that of the parent.  We
   describe the orientation of a tile by a matrix $M \in SO(3)$, the
   columns of which are unit vectors pointing along fixed features of
   the tile.  There exist fixed matrices $g_1,\ldots,g_8 \in SO(3)$
   such that the orientation of the $i$-th daughter is $M g_i$, where
   the orientation of the parent is $M$.  After $2$ subdivisions, the
   orientations of the granddaughters are of the form $M g_ig_j$.
   After $n$ subdivisions the orientations of the descendant tiles are
   equal to $M$ times words of length $n$ in the generators
   $\TBra{g_i}$.

\vbox{
\centerline{\epsfysize=2truein\epsfbox{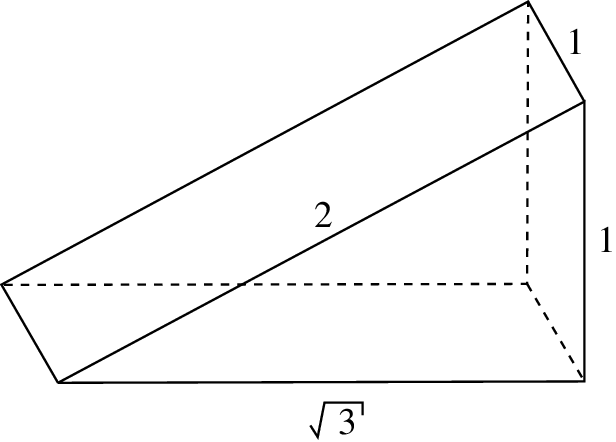}}
\smallskip
\centerline{\bf Figure 2a: A QQ tile.}
\smallskip
\smallskip
\centerline{\epsfysize=2.5truein\epsfbox{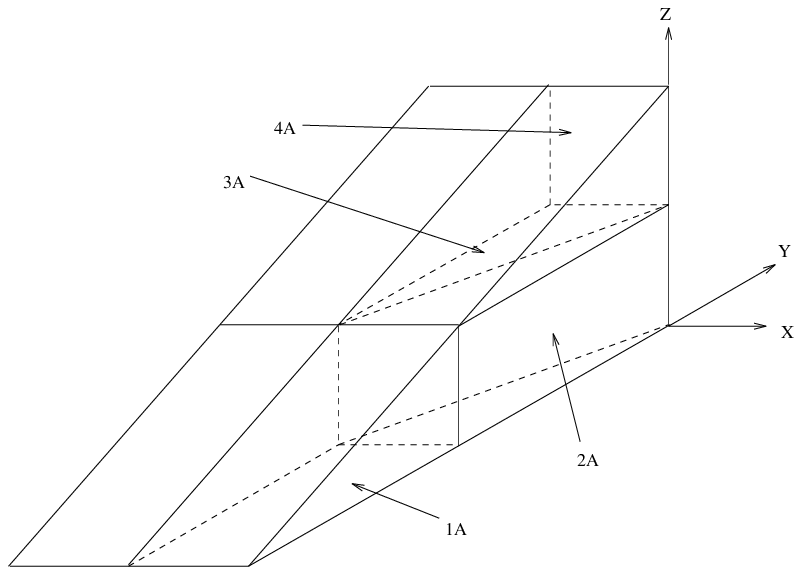}}
\smallskip
\centerline{\bf Figure 2b: The front four daughter tiles.}
\smallskip
\smallskip
\centerline{\epsfysize=2.5truein\epsfbox{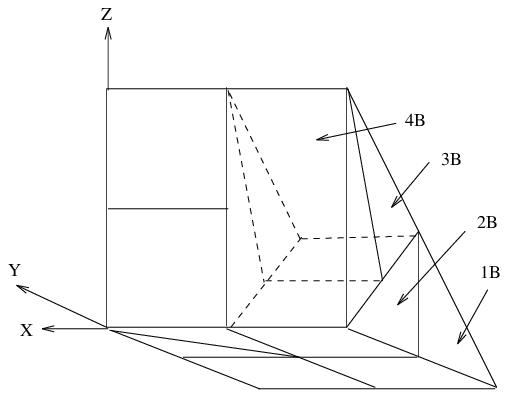}}
\smallskip
\centerline{\bf Figure 2c: The back four daughter tiles.}
}
\bigskip

   We now see why 2-dimensional tilings have extremely slow growth.
   $SO(2)$ is abelian, so the number of words of length $n$ in a fixed
   set of generators has only 
   polynomial growth in $n$, while the sample size of a substitution tiling
   is exponential in the number of subdivisions.  In other words, the
   number of orientations is logarithmic in the sample size.  $SO(3)$ is
   nonabelian, however, so the number of distinct words of length $n$ grows
   exponentially with $n$, hence as a power of the sample size.  
   The power, however, need not be large.  For the
   quaquaversal tiling we will prove
   
   \begin{Theorem}{T1}{This was originally theorem 1}
      Let $n \ge 2$. 

      If $n = 2k$ is even, then the number of 
      distinct orientations in the
      $n$-th subdivision of a tile is 
      $24(2^k-1)=24(N^{1/6}-1)$, where $N = 8^{n}$
      is the total number of tiles.  

      If $n = 2k+1$ is odd, the number of
      distinct orientations in the $n$-th subdivision is
      $34 \times 2^k - 24= (34/\sqrt{2})N^{1/6} -24$.
   \end{Theorem}
   
   By Theorem 1.1, a macroscopic sample of, say, $10^{24}$ tiles would
   appear in over 200,000 different orientations.  The orientations do
   not, however, appear with equal frequency, and they are not evenly
   distributed in $SO(3)$.  In a sample of $8^{27} \approx 2.4 \times
   10^{24}$ tiles, a mere 104 orientations account for 69\% of the
   tiles, while in a sample of $8^{120} \approx 2.3 \times 10^{108}$
   tiles, 1000 orientations account for slightly over half the tiles.
   The asymptotic behavior is given by:
   
   \begin{Theorem}{T2}{This was originally theorem 2}
      Let $\rho > 1/28$, and let $F_\alpha(n)$
      be the number of distinct orientations needed to account for a fraction
      $\alpha$ of the $8^n$ tiles in the $n$-th subdivision.  
      For any fixed $\alpha$, 
      \begin{eqnarray*}
        \lim_{n\to\infty}2^{-\rho n} F_\alpha(n) & = & 0.
      \end{eqnarray*}
   \end{Theorem}
   
   That is, $2^{-n/28}F_\alpha(n)$ is asymptotically smaller than any
   exponential in $n$ (i.e. any power of the volume), so we speak of
   $F_\alpha(n)$ as growing essentially as $2^{n/28}$, or as the 84th
   root of the volume.
   
      Finally, we consider the rate at which the orientations approach a
   uniform distribution on $SO(3)$.  Let $f(M)$ be a real-valued function
   of $M \in SO(3)$, and let $f_n(M)$ be the average value of $f$ on the
   $8^n$ orientations of the $n$-th order descendants of a tile of
   orientation $M$.  If $f$ is continuous, then $f_n(M)$ must converge to
   $\int_{SO(3)}f d\mu$ for every $M$, where $d\mu$ is Haar measure on
   $SO(3)$.  However, this convergence may be quite slow:
   
   \begin{Theorem}{T3}{This was originally theorem 3}
      There exists a smooth 
      function $f: SO(3) \to \Reals$ and a point $M \in SO(3)$ such that
      $|f_n(M)- \int_{SO(3)}f d\mu| > (0.9938)^{n}$ for every $n$.
   \end{Theorem}
   
   We believe this result to be close to sharp:

   \begin{Conjecture}{C1}{} For any continuous function $f: SO(3) \to \Reals$ 
    	and any point $M \in SO(3)$, there exists a constant $C$ such 
	that $|f_n(M) - \int_{SO(3)} f d\mu| < C 2^{-n/112} 
	\approx C (0.9938303)^n$ for every $n$.
   \end{Conjecture}

   In other words, it appears that functions do converge to their
   average values exponentially in the volume, with the exponents
   bounded away from 0, but there are exponents about as small
   as 1/336. As in all linear systems of exponential decay, the
   long-time behavior is dominated by the smallest exponents, so we
   expect the anisotropy of a generic function on $SO(3)$ to decay roughly as
   $N^{-1/336}$ for sufficiently large $N$.
   
   The outline of the paper is as follows.  In \myref{Section}{group} 
   we examine the set of
   orientations that can arise from successive subdivision of a single tile.
   This set turns out to be a subgroup, denoted $G(6,4)$, of $SO(3)$.
   The analysis of this section is a special case of some general results
   of \cite{RS1}.  We establish a canonical form for an element of $G(6,4)$
   as a word in two generators, 
   and provide an explicit list of the orientations in the $n$-th
   subdivision, thereby proving \myref{Theorem}{T1}.  We also develop a 
   measure, called ``size'', of the complexity of an element of $G(6,4)$.

   In \myref{Section}{PD} we consider the populations of the various
   orientations of the descendants of a given tile as a function of
   $n$, the number of subdivisions.  We show that the ``size'' of the
   orientation of a randomly chosen tile is described by a biased
   random walk (on the non-negative integers) with a reflecting
   barrier.  Although the maximum size is $n/2$, the distribution of
   sizes is approximately Gaussian with mean $n/28 + O(1)$.  Since
   there are $O(2^{n/28})$ words of size $n/28$ or less,
   \myref{Theorem}{T2} follows.

   \myref{Theorem}{T2} is an asymptotic statement.  To obtain results
   for specific values of $n$ we numerically implement the 
   random walk for $n$ up to 200.  This establishes the claims 
   previously made about samples of size $8^{27}$ and $8^{120}$.

   In \myref{Section}{Apperg}, we consider the rate at which the
   distribution of orientations approaches uniformity on $SO(3)$.
   There is a linear map $\call$ on $L^2(SO(3))$ such that $f_n =
   \call^n f$.  $\call$ is similar to the Hecke operators considered
   in \cite{LPS1, LPS2}, except that those Hecke operators were Hermitian,
   while $\call$ is not.  Nonetheless, the spectrum of $\call$ appears
   to be entirely real.  If $f_n$ is to approach a limiting function
   exponentially, the base of the exponent (for generic $f$) is the
   largest eigenvalue of $\call$ that is less than 1. The operator
   $\call$ is the direct sum of finite dimensional blocks, each block
   corresponding to an irreducible representation of $SO(3)$. By
   numerically diagonalizing blocks corresponding to the first 300
   representations (blocks with total dimension over 36,000,000), we
   find a large number of eigenvalues over 0.993, with the largest
   being approximately 0.99381, and conjecture that the eigenvalues
   are bounded above by $2^{-1/112}\approx 0.9938303$.

\mysection{The group $G(6,4)$}{group}

   Let $\Rx^\theta$ denote a rotation about the $x$ axis by an angle
   $\theta$.  The group generated by $\Rx^{2\pi/p}$ and $\Ry^{2\pi/q}$ is
   denoted $G(p,q)$.   (For a discussion of $G(p,q)$ for general $p$ and
   $q$, and for further generalizations, see \cite{RS1,RS2}.)  
   The set of orientations in the QQ tiling is
   precisely $G(6,4)$, generated by $S := \Ry^{\pi/2}$ and $T:=
   \Rx^{\pi/3}$.  Indeed, the orientations $g_1,\ldots,g_8$ of the eight
   daughters of a tile in standard position all are words in $S$ and $T$,
   and it is not hard to get $S$ and $T$ themselves as products of the
   $g_i$'s.  To understand these orientations, therefore, we must understand
   $G(6,4)$.
   
   A word $\alpha$ in $S$ and $T$ is said to be in {\it canonical form} if
   \begin{eqnarray}
      \alpha & = & S^{a_0} T^{b_0}, 
      \mylabel{Equation}{E1}
   \end{eqnarray}
   with $a_0 \in \{0,1,2,3\}$ and $b_0 \in \{ 0,3\}$, or if 
   \begin{eqnarray} 
      \alpha & = & W S T^{b_1}  S T^{b_2} \cdots S T^{b_n} E, 
      \mylabel{Equation}{E2}
   \end{eqnarray}
   with $n \ge 1$, with $W$ of the form \myeref{E1}, with each $b_i=2$ or 4, and
   with $E=1$ or $S$.  We shall soon see that each element of $G(6,4)$
   can be expressed by a unique canonical form.  
   The {\it size} of an element of $G(6,4)$ with canonical form \myeref{E2} is 
   defined to be $n$. (Elements of the form \myeref{E1} have size zero).
   
   To understand the group $G(6,4)$, we represent $S$ and $T$ by explicit
   $3 \times 3$ matrices.  Relative to the standard basis $\{e_1, e_2,e_3\}$
   of $\Reals^3$, the matrices of $S$ and $T$ are
   \begin{eqnarray} 
      S \; = \; \pmatrix{ 0 & 0 &  1 \cr 
                          0 & 1 &  0 \cr 
                         -1 & 0 &  0} 
      & \hbox{and} &
      T \; = \; \pmatrix{1 & 0          & 0 \cr 
                         0 & 1/2        & -\sqrt{3}/2 \cr 
                         0 & \sqrt{3}/2 & 1/2}.
      \mylabel{Equation}{E3}
   \end{eqnarray}
   The factors of $\sqrt{3}$ are inconvenient to work with.  We eliminate them
   by working in the basis $\{e_1,\sqrt{3}e_2, e_3\}$, relative to which the
   matrices of $S$ and $T$ are:
   \begin{eqnarray}
      S \; = \; \pmatrix{ 0 & 0 &  1 \cr 
                          0 & 1 &  0 \cr 
                         -1 & 0 &  0}; 
        & \qquad  &
      T \; = \; \pmatrix{1 & 0 & 0 \cr 0 & 1/2 & -1/2 \cr 0& 3/2 & 1/2}.
      \mylabel{Equation}{E4}
   \end{eqnarray} 
   Three other useful matrices are
   \begin{eqnarray}
      T^3 \; = \; \pmatrix{1 &  0 &  0 \cr 
                           0 & -1 &  0 \cr 
                           0 &  0 & -1}; 
        & \qquad  &
      ST^{\pm 2} \; = \; \pmatrix{ 0 & \pm 3/2 & -   1/2 \cr 
                                   0 & -   1/2 & \mp 1/2 \cr 
                                  -1 &       0 &     0}. 
      \mylabel{Equation}{E5}
   \end{eqnarray} 
   
   \begin{Theorem}{T4}{This was originally theorem 4} Every element $g
      \in G(6,4)$ can be expressed in canonical form, and the
      canonical form is unique.  If $g$ has size $n$, then, 
      in the representation (2.4), $2^n g$ is an integer matrix with
      at least one odd matrix element.  There exist exactly 8 group
      elements of size 0 and $16 \times 2^n$ elements of size $n>0$.
      \end{Theorem}
   
   \begin{proof}
      Note that $2ST^2$ and $2ST^4$ are matrices with 
      integer entries of the form
      \begin{eqnarray}
         \pmatrix{even & odd  & odd \cr 
                  even & odd  & odd \cr 
                  even & even & even},
         \mylabel{Equation}{E6}
      \end{eqnarray}
      and that the product of two or more matrices of the form 
      given in \myeref{E6} is also of
      the form given in \myeref{E6}.   
      If $g$ can be expressed
      in canonical form \myeref{E6}, then
      \begin{eqnarray}
         2^n g & = & W (2S T^{b_1})(2  S T^{b_2}) \cdots (2S T^{b_n}) E 
                     \nonumber \\
               & = &  W \myspecref{E6} \myspecref{E6} \cdots \myspecref{E6}E 
                     \nonumber \\
               & = & W \myspecref{E6} E.
      \mylabel{Equation}{E7}
      \end{eqnarray}
      Since $S$ is a permutation matrix (up to sign), 
      and $T^3$ is the identity (up to sign), $2^n g$
      is an integer matrix with four odd matrix elements and 
      five even matrix elements. If $g$ is of canonical form \myeref{E1}, then
      $2^n g = 2^0 g=g$ is an integer matrix with three odd matrix
      elements. This proves the second statement of the theorem, and also
      shows that the identity can be put in canonical form in only one
      way, namely $1=S^0T^0$.
      
      To see that every element of $G(6,4)$ can be put in canonical form,
      begin with an arbitrary word 
      \begin{eqnarray} 
         \alpha & = & S^{a_1} T^{b_1} \cdots S^{a_n} T^{b_n} 
         \mylabel{Equation}{E8}
      \end{eqnarray} 
      in $S$ and $T$.  If any of the $a_i$'s is even, or any of the $b_i$'s is
      a multiple of 3, we can use the relations 
      \begin{eqnarray} 
         ST^3 & = & T^3S^{-1} 
         \mylabel{Equation}{E9}
      \end{eqnarray} 
      and 
      \begin{eqnarray}
         S^2 T & = &T^{-1}S^2 
         \mylabel{Equation}{E10}
      \end{eqnarray} 
      to shorten the word.  So we can assume that all the $a_i$'s
      (except possibly $a_1$) are odd, and that none of the $b_i$'s 
      (except possibly $b_1$) are multiples of 3. 
      We then use \myeref{E9} to extract powers of $T^3$
      and move them leftwards, changing $T^1$ to $T^4$ and 
      changing $T^5$ to $T^2$.
      We then use \myeref{E10} to extract powers of $S^2$ 
      and move them leftwards, 
      changing $S^3$ to $S$.  
      In the process, we may change some $T^2$'s to $T^4$'s 
      and vice-versa, but that does not affect the form \myeref{E2}.  
      Finally, if $b_1$ is not divisible by 3, 
      we use \myeref{E9} to rewrite $S^{a_1}T^{b_1}$
      as $S^{a_0} T^{3b_0} S T^{b_1'}$ with $b_1'=2$ or $4$.
      
      To see that a canonical form is unique, suppose $g$ and $g'$ are distinct
      canonical forms for the same group element.  
      Using the relations \myeref{E9} and \myeref{E10},
      we could then shorten the word $g' g^{-1}$ to a nontrivial word in 
      canonical form, which would have to represent the identity.  But we have
      already seen that there
      are no nontrivial canonical forms for the identity.  
      
      Since group elements are in 1-1 correspondence with canonical forms,
      we can count group elements of any given size.  There are 8 choices 
      for $W$, 2 choices for each $b_i$, and 2 choices for $E$, for a total
      of $16\times 2^n$ elements of size $n>0$.  For size 0 there are
      only 8 choices, 4 for $a_0$ and 2 for $b_0$. 
   \end{proof}
   
   We now have the machinery to understand the orientations of tiles
   in the $n$-fold subdivision of a single quaquaversal tile.  In a
   single subdivision, the daughters are described by group elements
   $\{g_1,\ldots,g_8\}$ with 

\begin{equation} \begin{array}{rcccl} g_1
   & = & \Rx^{0} & = & 1 \\ g_2 & = & \Ry^{3\pi/2}\Rz^\pi & = & ST^3
   \\ g_3 & = & \Ry^{\pi/2} & = & S \\ g_4 & = & \Rx^{0} & = & 1 \\
   g_5 & = & \Rx^{0} & = & 1 \\ g_6 & = & \Rz^{\pi} & = & S^2T^3 \\
   g_7 & = & \Rx^{4\pi/3}\Ry^\pi & = & T^4S^2 \; = \; S^2T^2 \\ g_8 &
   = & \Rx^{4\pi/3} & = & T^4 \end{array} \mylabel{Equation}{E11}
\end{equation} 

Note that $T=g_7g_6$, so the semigroup generated by the $g_i$'s is
precisely the group generated by $S$ and $T$.  Also note that only two
of the $g_i$'s, namely $g_7$ and $g_8$, have size 1, while the rest
have size zero. Since $g_7^2$, $g_8^2$, $g_7g_8$ and $g_8g_7$ all have
size 1 or less, the orientations in the 2nd subdivision have size at
most 1 and the orientations in the $2k$-th subdivision have size at
most $k$.  This shows that the number of possible orientations in the
$n$-th subdivision is bounded by a constant multiple of $2^{n/2}$.

However, we can be more precise.  The following theorem implies Theorem 1.1:

   \begin{Theorem}{T1new}{This implies theorem 1} The orientations
      appearing in subdivisions of a single tile (of standard
      orientation) are as follows.  Let $P_m$ be the set of words of length
      $m$ in the two generators $ST^2$ and $ST^4$.  For $k>0$, 
      the orientations of the $2k$-th subdivision are:
\begin{enumerate} 
	\item $T^b P_{m} S^a$, with no restriction on
      $a$ or $b$, and with $m$ ranging from 0 to $k-2$, 
	\item $T^b
      P_{k-1} S^a$, with $b=0$, 2, 3, or 5 and $a$ arbitrary, 
	\item
      $T^b P_{k-1} S^a$ with $b=1$ or 4 and $a=0$ or 2, 
	\item $ST^2
      P_{k-1} S^a$, with $a=0$ or 2, and \item $T^3 (ST^4) P_{k-1}
      S^a$, with $a=0$ or 2.  
\end{enumerate}
      
      For $k>0$, the orientations in the 
      $(2k+1)$st subdivision are
      \begin{enumerate}
         \item $T^b P_{m} S^a$, with no restriction on $a$ or $b$ and $m$ ranging
               from $0$ to $k-1$,
         \item $T^b P_{k} S^a$ , with $b=0$, 2, 3 or 5 and $a=0$ or 2,
         \item $ST^2 P_{k-1} S^a$, with $a=1$ or 3, and 
         \item $T^3 ST^4 P_{k-1} S^a$, with $a=1$ or 3.
      \end{enumerate}
\end{Theorem}

\begin{proof}
      It is not hard to see that, in each case, the sets are disjoint
      and their cardinalities add up to the total indicated in Theorem
      1.1.  To complete the proof we must check i) that the claimed
      orientations for 2 subdivisions are correct, ii) that the
      claimed orientations for $2k$ subdivisions times the generators
      $g_1,\ldots,g_8$ give precisely the claimed orientations for
      $2k+1$ subdivisions, and iii) that the claimed orientations for
      $2k+1$ subdivisions times the generators $g_1,\ldots,g_8$ give
      precisely the claimed orientations for $2k+2$ subdivisions.  We
      leave this straightforward (but quite tedious) computation as an
      exercise for the highly motivated reader.  
\end{proof}
   
\mysection{Population dynamics}{PD}
   
   In \myref{Section}{group}, 
   we counted the number of distinct orientations of tiles in the
   $n$-th subdivision of a QQ tile, which equals the number of
   algebraically distinct words of length $n$ in the generators 
   $g_1,\ldots g_8$.  In this section we consider the distribution of tiles
   among the various orientations.  Equivalently, we count repetitions in 
   the words of length $n$ in the $g_i$'s.  

   We do this by modeling the subdivision process as a random walk on
   $G(6,4)$.  If at some stage we are at the group element $M$, then
   at the next stage we can go to $Mg_1, Mg_2, \ldots, Mg_8$, each
   with probability 1/8.  The distribution of orientations in the
   $n$-th subdivision of a tile (with standard orientation) is
   precisely the $n$-th iterate of this random walk, starting at the
   origin.  The following theorem implies \myref{Theorem}{T2}.
   
   \begin{Theorem}{T5}{This was originally theorem 5}
      Consider the random walk starting at the origin,
      and let $P(k,n)$ be the probability that the $n$-th iteration yields 
      a group element of size $k$ or greater.  For any $\rho>1/28$,
      $\lim_{n \to \infty} P(n\rho,n)=0$.
   \end{Theorem}
   
   \begin{proof}
      
      We divide $G(6,4)$ into sets according to size and the pattern
      at the end of the canonical form.  The random walk on
      $G(6,4)$ then induces a random walk on the space of sets, and we
      show that the bias in this random walk causes the average size
      to increase by 1/28 each turn.  The result then follows from the
      central limit theorem.

      Let $R(k,ST^2)$ be the set of
      group elements of size $k$ whose canonical forms end with
      $ST^2$.  $R(k,ST^4)$, $R(k,T^2S)$, and $R(k,T^4S)$ are defined
      similarly. $R(0)$ is the set of elements of size 0.
      
      If $M \in R(k,ST^2)$, then $Mg_1$, $Mg_4$ and $Mg_5$ are equal
      to $M$, and so are in $R(k,ST^2)$.  $Mg_2$ is in $R(k,T^4S)$,
      $Mg_3$ is in $R(k,T^2S)$, and $Mg_6$ is in $R(k,ST^4)$.  $Mg_7$
      may be in $R(k-1,T^2S)$ or $R(k-1,T^4S)$, depending on whether
      $M$ ends with $ST^2ST^2$ or $ST^4ST^2$.  However, if $Mg_7\in
      R(k-1,T^2S)$, then $Mg_8 \in R(k-1,T^4S)$, and vice-versa.  In
      either case, starting in $R(k,ST^2)$, there is a 3/8 probability
      of staying in $R(k,ST^2)$ and a 1/8 probabilty for each of the
      destinations $R(k,ST^4)$, $R(k,T^2S)$, $R(k,T^4S)$,
      $R(k-1,T^2S)$ and $R(k-1,T^4S)$.

      This analysis, and the corresponding analysis for $R(k,ST^4)$,
      $R(k,T^2S)$ and $R(k,T^4S)$, is summarized in the diagram

\begin{center}
\unitlength=1mm
\begin{picture}(130,40)(0,0)
\put(5,31){\makebox(0,0){$R(k\!-\!1,T^2S)$}}
\put(45,31){\makebox(0,0){$R(k,ST^2)$}}
\put(85,31){\makebox(0,0){$R(k,T^2S)$}}
\put(125,31){\makebox(0,0){$R(k\!+\!1,ST^2)$}}
\put(5,5){\makebox(0,0){$R(k\!-\!1,T^4S)$}}
\put(45,5){\makebox(0,0){$R(k,ST^4)$}}
\put(85,5){\makebox(0,0){$R(k,T^4S)$}}
\put(125,5){\makebox(0,0){$R(k\!+\!1,ST^4)$}}
\put(20,31){\vector(1,0){12}}		
\put(32,30){\vector(-1,0){12}}		
\put(60,31){\vector(1,0){12}}		
\put(108,30){\vector(-1,0){12}}		
\put(96,31){\vector(1,0){12}}		
\put(20,5){\vector(1,0){12}}		
\put(60,5){\vector(1,0){12}}		
\put(96,5){\vector(1,0){12}}		
\put(18,8){\vector(1,1){20}}		
\put(35,28){\vector(-1,-1){18}}		
\put(12,28){\vector(4,-3){24}}		
\put(51,28){\vector(4,-3){24}}		
\put(74,8){\vector(-4,3){24}}		
\put(44,8){\vector(0,0){20}}		
\put(46,28){\vector(0,-1){20}}		
\put(56,8){\vector(1,1){20}}		
\put(73,28){\vector(-1,-1){18}}		
\put(94,8){\vector(1,1){20}}		
\put(111,28){\vector(-1,-1){18}}	
\put(91,28){\vector(4,-3){24}}		
\put(45,35){\oval(4,6)[t]}		
\put(47,35){\vector(0,-1){1}}		
\put(40,39){$\scriptstyle g_1,g_4,g_5$}		
\put(85,35){\oval(4,6)[t]}		
\put(87,35){\vector(0,-1){1}}		
\put(78,39){$\scriptstyle g_1,g_4,g_5,g_6$}	
\put(25,33){$\scriptstyle g_7$}				
\put(22,28){$\scriptstyle g_7\mbox{ \rm \small or }g_8$} 	
\put(63,33){$\scriptstyle g_3$}				
\put(100,33){$\scriptstyle g_7$}			
\put(97,28){$\scriptstyle g_7\mbox{ \rm \small or }g_8$} 
\put(10,14){$\scriptstyle g_8\mbox{ \rm \small or }g_7$} 
\put(86,14){$\scriptstyle g_8\mbox{ \rm \small or }g_7$} 
\put(50,20){$\scriptstyle g_2,g_3$} 			
\put(52,14){$\scriptstyle g_2,g_3$} 			
\put(40,20){$\scriptstyle g_6,$}			
\put(40,17){$\scriptstyle g_8$}				
\put(47,17){$\scriptstyle g_6$}				
\put(36,24){$\scriptstyle g_7$}				
\put(74,24){$\scriptstyle g_2$}				
\put(113,24){$\scriptstyle g_7$}			
\put(31,14){$\scriptstyle g_8$}				
\put(70,14){$\scriptstyle g_2$}				
\put(110,14){$\scriptstyle g_8$}			
\put(25,3){$\scriptstyle g_8$}				
\put(63,3){$\scriptstyle g_3$}				
\put(100,3){$\scriptstyle g_8$}				
\put(45,2){\oval(4,6)[b]}		
\put(47,2){\vector(0,0){1}}		
\put(40,-3){$\scriptstyle g_1,g_4,g_5,g_7$}	
\put(85,2){\oval(4,6)[b]}		
\put(87,2){\vector(0,0){1}}		
\put(78,-3){$\scriptstyle g_1,g_4,g_5,g_6$}	
\end{picture}
\end{center}

This diagram continues to the right indefinitely, but to the left it ends
at $k=0$ with

\begin{center}
\unitlength=1mm
\begin{picture}(50,40)(0,0)
\put(15,20){$R(0)$}
\put(40,40){$R(1,ST^2)$}
\put(40,1){$R(1,ST^4)$}
\put(24,22){\vector(1,1){16}}            
\put(39,39){\vector(-1,-1){15}}         
\put(24,20){\vector(4,-3){18}}          
\put(12,21){\oval(6,5)[l]}              
\put(12,18.5){\vector(1,0){1}}           
\put(7,16.0){$\scriptstyle g_1,g_2,g_3,$}         
\put(7,13.5){$\scriptstyle g_4,g_5,g_6$}         
\put(26,35){$\scriptstyle g_7,g_8$} 
\put(38,33){$\scriptstyle g_7$} 
\put(32,10){$\scriptstyle g_8$}
\end{picture}
\end{center}

      Since the long-time behaviors of positively biased random walks
      are independent of the boundary conditions at $k=0$, the
      reflecting barrier at $k=0$ does not affect the theorem.  To
      establish the theorem, we need only calculate the bias in the
      random walk away from $k=0$ and show it equals 1/28.
      
      We first compute the distribution of $ST^2$, $ST^4$, $T^2S$ and
      $T^4S$ endings.  Summing over $k$, we find the transition matrix
      for the four classes of group elements is 
\begin{equation} {1
      \over 8} \pmatrix{3 & 2 & 1 & 3 \cr 1 & 4 & 3 & 1 \cr 2 & 1 & 4
      & 0 \cr 2 & 1 & 0 & 4}.  \mylabel{Equation}{E12} 
\end{equation}
      The limiting distribution is the eigenvector with eigenvalue 1,
      namely 
\begin{equation} 
{1 \over 14} \pmatrix{4 \cr 4 \cr 3 \cr3}.  \mylabel{Equation}{E13} 
\end{equation} 
      That is, asymptotically $4/14$ of the tiles have orientations
      whose canonical forms end with $ST^2$, 4/14 end with
      $ST^4$, 3/14 end with $T^2S$ and 3/14 end with $T^4S$.
      
      On each iteration, an $ST^2$ state has a $2/8$ chance of decreasing
      $k$ (and none of increasing), $ST^4$ states have no chance of either
      increasing or decreasing $k$, while $T^2S$ and $T^4S$ states each have 
      a 2/8 chance of increasing $k$ and no chance of decreasing.  Weighing
      these chances by the $(4/14,4/14,3/14,3/14)$ limiting distribution, 
      we see that at each step in the random walk
      there is a 3/28 chance of increasing $k$ and a 2/28 chance of 
      decreasing $k$, for a net bias of 1/28.  
   \end{proof}
   
   \myref{Theorem}{T5} is an asymptotic result that does not indicate
   have many iterations are needed to reach the asymptotic regime.
   Indeed, the bias in the random walk is so small that, for moderate
   numbers of iterations, one would expect diffusion and the
   reflecting barrier at $k=0$ to be quite significant. To measure
   this transient behavior we rely on the following numerical
   implementation, in MATLAB, of the biased random walk.  Here {\tt
   zz(n)} is the fraction of orientations of size zero in the $n$-th
   generation, {\tt st2(k,n)} (resp. {\tt st4(k,n), t2s(k,n),
   t4s(k,n)}) is the fraction in $R(ST^2,k)$ (resp. $R(ST^4,k)$,
   $R(T^2S,k)$, $R(T^4S,k)$) in the $n$-th generation, and {\tt
   tot(k+1,n)} is the fraction of size $k$ in the $n$-th generation.
   
{\tt \noindent \% Initialize matrices to zero \hfill\break
  for n=1:Nmax \hfill\break 
\quad\quad  zz(n)=0; \hfill\break 
\quad\quad  for k=1:(2+Nmax/2) \hfill\break 
\quad\quad\quad\quad  st2(k,n)=0; st4(k,n)=0; t2s(k,n)=0; t4s(k,n)=0; tot(k,n)=0;
\hfill\break end; 
end; 

\noindent \% Set up distribution for n=1 and iterate to n=Nmax. 
\hfill\break
zz(1)=6/8; st2(1,1)=1/8; st4(1,1)=1/8; tot(1,1)=6/8; tot(2,1)=2/8; \hfill\break
for n=2:Nmax \hfill\break
zz(n)=(6*zz(n-1)+2*st2(1,n-1))/8; \hfill\break 
st2(1,n)=(zz(n-1)+3*st2(1,n-1)+2*st4(1,n-1)+2*t4s(1,n-1))/8; 
\hfill\break
st4(1,n)=(zz(n-1) + st2(1,n-1)+4*st4(1,n-1)+2*t2s(1,n-1))/8;\hfill\break
t2s(1,n)=(st2(1,n-1)+st4(1,n-1)+4*t2s(1,n-1)+st2(2,n-1))/8;\hfill\break
t4s(1,n)=(st2(1,n-1)+st4(1,n-1)+4*t4s(1,n-1)+st2(2,n-1))/8;\hfill\break
tot(1,n)=zz(n);  tot(2,n)=st2(1,n)+st4(1,n)+t2s(1,n)+t4s(1,n);\hfill\break
for k=2:(1+Nmax/2) \hfill\break
st2(k,n)=(t2s(k-1,n-1)+t4s(k-1,n-1)+3*st2(k,n-1)+2*st4(k,n-1)+2*t4s(k,n-1))/8;
st4(k,n)=(t2s(k-1,n-1)+t4s(k-1,n-1) + st2(k,n-1)+4*st4(k,n-1)+2*t2s(k,n-1))/8;
t2s(k,n)=(st2(k,n-1)+st4(k,n-1)+4*t2s(k,n-1)+st2(k+1,n-1))/8; \hfill\break
t4s(k,n)=(st2(k,n-1)+st4(k,n-1)+4*t4s(k,n-1)+st2(k+1,n-1))/8; \hfill\break
tot(k+1,n)=st2(k,n)+st4(k,n)+t2s(k,n)+t4s(k,n); \hfill\break
end; end; \hfill\break
tot \% Output results.}

Running this program with {\tt Nmax}=200, one sees that the median orientation
size is
$$ \hbox{Median size } = \cases{0 & if $n \le 2$; \cr
1 & if $2 < n \le 20$; \cr
2 & if $20 < n \le 42$; \cr
3 & if $42 < n \le 67$; \cr
4 & if $67 < n \le 93$; \cr
5 & if $93 < n \le 120$; \cr
6 & if $120 < n \le 147$; \cr
7 & if $147 < n \le 174$; \cr
8 & if $174 < n \le 200$.}
$$
Note that by Theorem 2.6 there are exactly 8 orientations of size
zero, 40 of size at most 1, 104 of size at most 2, 232 of size at most
3, 488 of size at most 4, 1000 of size at most 5, 2024 of size at most
6, 4072 of size at most 7 and 8168 of size at most 8.  Thus for
$n=27$, just 104 orientations account for the majority of the tiles,
while for $n=120$, a mere 1000 orientations account for the majority
of the tiles.

Table 1 shows the distribution of sizes for three samples, a
microscopic sample with $8^5 = 32768$ tiles, a macroscopic sample with
$8^{27} \approx 2.42 \times 10^{24}$ tiles (about 4 moles), and an
impossibly large sample of $8^{120} \approx 2.35 \times 10^{108}$
tiles. Table 2 shows similar results for $n=10$, 20, 30, 40 and 50. 

\begin{table}\vbox{\hfil
\begin{tabular}{|c|r|r|r|} \hline
Size & \hfil $n=8$ \hfil &  \hfil $n=27$ \hfil &\hfil $n=120$ \hfil \\ \hline
0 &     0.2588  &       0.0971   &      0.0145   \\
1 &     0.5531  &      0.3023  &     0.0524   \\
2 &     0.1719  &       0.2931   &     0.0788   \\
3 &     0.0159  &       0.1903  &    0.1038   \\
4 &     0.0003  &       0.0847   &     0.1221   \\
5 & 0 &       0.0260   &     0.1299   \\
6 & 0 &       0.0055  &    0.1256   \\
7 & 0 &       0.0008  &     0.1110   \\
8 & 0 &       0.0001 &       0.0897   \\
9 & 0 & 0&    0.0665   \\
10& 0 & 0&    0.0453   \\
11 & 0 & 0 &    0.0283   \\
12 & 0 & 0 &     0.0163   \\
13 & 0 & 0 &     0.0086   \\
14 & 0 & 0 &     0.0042   \\
15 & 0 & 0 &    0.0019   \\
16 & 0 & 0 &    0.0008   \\
17 & 0 & 0 &    0.0003   \\
18 & 0 & 0 &    0.0001  \\ 
19 & 0 & 0 &   0.0000 \\   \hline
\end{tabular}\hfil}
\caption{Distributions of Sizes in Various Samples.}
\end{table}

\begin{table}\vbox{\hfil
\begin{tabular}{|c|r|r|r|r|r|} \hline
Size & \hfil $n=10$ \hfil &  \hfil $n=20$ \hfil &\hfil $n=30$ \hfil 
& \hfil $n=40$ \hfil & \hfil $n=50$ \hfil
\\ \hline
0 & 0.2199 &  0.1272 &   0.0878   &  0.0654   &     0.0508   \\
1 & 0.5230   & 0.3738  & 0.2780   &  0.2154   &    0.1718   \\
2 & 0.2199   & 0.3023  & 0.2842   &  0.2488   &     0.2140   \\
3 & 0.0353   & 0.1457  & 0.2013   &  0.2149   &    0.2089   \\
4 & 0.0019   & 0.0426  & 0.1011   &   0.1422   &    0.1635   \\
5 & 0.0000   & 0.0075  &  0.0363   &   0.0728   &     0.1038   \\
6 & 0 &  0.0008  & 0.0093   &   0.0290   &     0.0537   \\
7 & 0 &  0.0000  & 0.0017   &  0.0090   &     0.0227   \\
8 & 0 &  0.0000  & 0.0002   &  0.0021   &     0.0079   \\
9 & 0 &  0.0000  & 0.0000   &   0.0004   &    0.0022   \\
10 &0 &  0.0000  & 0.0000   &  0.0001   &     0.0005   \\ \hline
\end{tabular}\hfil}
\caption{Additional Distributions of Sizes.}
\end{table}
        
\mysection{Approach to the ergodic limit}{Apperg}
   
   Let $f(M)$ be a function on $SO(3)$ and let $f_n(M)$ be the average of
   $f$ over the $8^n$ tiles in the $n$-fold subdivision of a tile of orientation
   $M$.  Equivalently, 
   \begin{eqnarray*} 
      f_n(M) & = & 8^{-n}\sum_{i_1=1}^8\sum_{i_2=1}^8\cdots \sum_{i_n=1}^8
      f(Mg_{i_1}g_{i_2}\cdots g_{i_n}).  
      \mylabel{Equation}{E14}
   \end{eqnarray*} 
   As $n \to \infty$, $f_n(g)$ approaches the constant limit $\int_{SO(3)} f
   d\mu$, where $d \mu$ is Haar measure.  The question is how fast the
   (ergodic) limit is approached.
   
   One approach to the problem is to consider the linear operator 
   $\call: L^2(SO(3)) \to L^2(SO(3))$ defined by
   \begin{eqnarray} 
      \call f(M) & =  & {1\over 8} \sum_{i=1}^8 f(Mg_i). 
      \mylabel{Equation}{E15}
   \end{eqnarray} 
   We can expand an arbitrary function in eigenfunctions $\xi_i$ of $\call$,
   with eigenvalues $\lambda_i$:
   \begin{eqnarray}
      f & = & \sum_i a_i \xi_i, 
      \mylabel{Equation}{E16}
   \end{eqnarray}
   so that
   \begin{eqnarray}
      f_n & = & \sum_i a_i \lambda_i^n \xi_i. 
      \mylabel{Equation}{E17}
   \end{eqnarray}
   Of course, the constant function is an eigenfunction of $\call$ with 
   eigenvalue 1.  The rate at which $f_n$ approaches a constant function
   is determined by the second-largest eigenvalue of $\call$.  

The operator $\call$ is quite similar to the Hecke operators
considered in \cite{LPS1, LPS2}, except that the set $\{g_i\}$ is not
balanced, as the inverses of $g_3$ and $g_8$ are not in the set.  As a
result, $\call$ is not a self-adjoint operator on $L^2(SO(3))$, not
even with a weighted measure, and we know of no {\it a priori} reason
why the eigenvalues of $\call$ need be real.  However, the numerical
experiments described below indicate that the eigenvalues of $\call$
are indeed real.  We do not understand the reason for this.
   
   The natural (left and right) actions of $SO(3)$ on $L^2(SO(3))$,
   namely $(l,r): f(g) \to f(lgr)$, decomposes $L^2(SO(3))$ into
   irreducible representations of $SO(3)$.  Since $\call$ is a linear
   combination of right-translations, it acts separately on each
   representation.  Thus the infinite-dimensional job of diagonalizing
   $\call$ on all of $L^2$ breaks down into a sequence of
   finite-dimensional problems, namely diagonalizing $\call$ on each
   irreducible representation that appears in the decomposition of
   $L^2$.
   
   The irreducible representations of $SO(3)$ are standard.  The spin-$\ell$
   representation, of dimension $2\ell+1$, appears $2 \ell+1$ times in the
   decomposition of $L^2$.  In particular, each representation appears at least
   once.  For each representation $R_\ell$, we must diagonalize the $(2\ell+1)
   \times (2\ell+1)$ matrix
   \begin{eqnarray}
      \call_\ell & = & {1\over 8} \DBkt{R_\ell(g_1) + \cdots + R_\ell(g_8)}. 
      \mylabel{Equation}{18}
   \end{eqnarray}

   We have done this, using MATLAB, for $\ell$ up to 300, thereby
   diagonalizing $\call$ on a subspace of $L^2(SO(3))$ of total
   dimension greater than 36,000,000.  There are numerous eigenvalues
   over 0.993, but none over 0.994; the largest known eigenvalue is
   $\lambda \approx 0.99381$, appearing in the $\ell=258$
   representation. This proves Theorem 1.4, and suggests there is an
   upper bound to the eigenvalues slightly above 0.99381.  Since
   powers of 2 appear repeatedly in the analysis, and since Theorem
   1.2 suggests a time scale of 28 subdivisions, $2^{-1/(4 \times 28)}
   \approx 0.9938303$ is a fairly natural guess for that upper bound.

\begin{table}\vbox{\hfil
\begin{tabular}{|c|r||c|r|} \hline
$\ell$ & \hfil Eigenvalue \hfil & $\ell$ & \hfil Eigenvalue \hfil \\ \hline
  1 &               0.50000  &  11&	 0.90142   \\
  2 &               0.78785  & 12	&  0.96793  \\
  3  &  0.50000  &  13	&  0.94705 \\ 
  4  &  0.92693  &  14	&  0.98797 \\ 
  5  &  0.62500  &  15	&  0.90546 \\
  6 &	 0.94912 &   16	&  0.96437 \\ 
  7&	0.84548  &  17	&   0.94949  \\
  8& 0.98454   & 18	&   0.99048 \\ 
  9&  0.73649  & 19	&   0.93382  \\
  10&	  0.95848 &  20	&  0.98042  \\ \hline  
\end{tabular}\hfil}
\caption{Largest Eigenvalues for First 20 Representations}
\end{table}

\begin{table}\vbox{\hfil
\begin{tabular}{|c|r|} \hline
$\ell$ & \hfil Eigenvalue \hfil \\ \hline
  1 &                        0.50000 \\ 
  2 &                        0.78785  \\
  4 &                        0.92693 \\
  6 &                        0.94912 \\
  8 &                        0.98454 \\
  14 &                       0.98797 \\
  18 &                       0.99048 \\
  32 &                       0.99243 \\
  45 &                       0.99324 \\
  56 &                       0.99335 \\
  72 &                       0.99362 \\
  248 &                      0.99367  \\
  258 &                      0.99381  \\ \hline
\end{tabular}\hfil}
\caption{Record Eigenvalues} 
\end{table}

   Table 3 lists the largest eigenvalue for each of the first 20
   representations.  Table 4 lists the ``record'' eigenvalues, those
   larger than any eigenvalue corresponding to smaller $\ell$. The
   MATLAB program for finding the largest eigenvalues from the first
   {\tt Lmax} representations is

\noindent {\tt for L=1:Lmax \hfill\break
\indent \% First specify angular momentum matrices \hfill\break
\indent j1=zeros(L+L+1); j2=j1; j3=j2;\hfill\break 
\indent for m=1:(L+L)  \hfill\break
\indent \indent j1(m,m+1)=(L+L+1-m)*i/2; j1(m+1,m)=m*i/2;  \hfill\break
\indent \indent 
j2(m,m+1)=(L+L+1-m)/2; j2(m+1,m)=-m/2; j3(m,m)=(L+1-m)*i; \hfill\break
\indent end; 
\hfill\break
\indent j3(L+L+1,L+L+1)=-L*i; \hfill\break
\indent S = expm(pi*j2/2); T=expm(pi*j1/3); \% Exponentiate to get rotations
\hfill\break
\indent A=S*T\^{}3; B=S; C=S\^{}2*T\^{}3; D=S\^{}2*T\^{}2; E=T\^{}4;  
\hfill\break 
\indent F=(3*eye(L+L+1)+A+B+C+D+E)/8; G=eig(F); 
H(L,1)=L; H(L,2)=max(G);  \hfill\break
end; H}

\mysection{Conclusions and Speculations}{Conc}

Much of the interest in tilings stems from the fact that crystals and
quasicrystals are modeled by tilings in which the tiles appear in only
a finite number of orientations.  Is there a form
of matter that is modeled on aperiodic tilings with statistical
rotational symmetry, in particular on a substitution tiling such as
the quaquaversal?  If so, how would such a form of matter be
recognized?

This paper provides the tools needed to recognize such a substance. We
have computed the distribution of orientations in a sample of size
$N=8^n$ of the QQ tiling.  The exact same calculation gives the
distribution of orientations of 8-tile clusters in a sample of size
$8^{n+1}$, of 64-tile clusters in a sample of size $8^{n+2}$, and
so on. In particular, the angular distribution of the relative
positions of nearby tiles, and therefore the angular distribution of
the electron diffraction pattern, are governed by the results of this
paper.

To detect an asymptotically round quasicrystal, one must look at the
diffraction pattern of a finite sample and decompose it into spherical
harmonics. As long as the wavelengths in question are much shorter than
the size of the sample, the angular distribution of the diffraction pattern
will scale with sample size in a manner determined by the eigenvalues.
If a spherical harmonic is an eigenvector of $\call$ with eigenvalue 
$\lambda$, then the diffraction pattern of a sample of size $8N$ will 
exhibit this harmonic a factor of $\lambda$ less than that of 
a sample of size $N$.

As a result, some harmonics will be essentially absent from the 
diffraction pattern of a macroscopic sample; those correspond to
eigenfunctions of $\call$ with 
$\lambda^n \ll 1$.  Other modes, with $1-\lambda$ comparable to or
less than $1/n$, will appear. A macroscopic (say, $N=8^{27}$)
sample of the QQ tiling, for example, would have a diffraction pattern
with a small but probably detectable $\ell=4$ component, and with
essentially undamped $\ell=6$, $\ell=8$ and $\ell=10$ components.
There are relatively few eigenfunctions with such large eigenvalues
(even for $\ell=8$, all but two of the 17 eigenvalues are less than
0.65), so the diffraction pattern will be dominated by the $\ell=0$
(isotropic) component, with some corrections from harmonics with
$\ell=6, 8, 10$, etc. 
This result is much more isotropic than
that of a crystal or ordinary quasicrystal (for which many harmonics are
completely undamped, and the spectrum consists of discrete points), 
while much less isotropic than that of an amorphous substance such as a glass,
for which only the isotropic $\ell=0$ term is observable.

The radial dependence of the diffraction pattern also carries
important information.  As with other quasicrystals, and unlike
amorphous media, one expects asymptotically round quasicrystals to
have radially self-similar diffraction patterns.  In combination, a
self-similar radial pattern combined with a mildly anisotropic angular
pattern should indicate an asymptotically round quasicrystal.

We thank Charles Radin for helpful discussions and Margaret Combs for
her \TeX{}nical expertise.  B.D. thanks J.M. Linhart for her guidance
and support.  The work of L.S. is partially supported by NSF Grant
No. DMS-9626698 and Texas ARP Grant 003658-152.

\bibliographystyle{plain} 
\bibliography{sadun}
\end{document}

%% file: texdefs.tex
\newcommand{\myfilename}{texdefs}
\newcommand{\mylabel}[3][\myfilename]{\label{#1:#2:#3}}
\newcommand{\myref}[3][\myfilename]{\ref{#1:#2:#3}}
\newcommand{\myeref}[2][\myfilename]{(\ref{\myfilename:Equation:#2})}

\newcommand{\outlinenatural}[3]{}
\newcommand{\outlinepoint}[3]{}
\newcommand{\outlinesimple}[2]{}
\newcommand{\setsource}[4]{\def\myfilename{#1}}
\newcommand{\showdraft}{}

\makeatletter
\newcommand{\goodoutlinenatural}[3]{\protected@write\outlinefile{}{\string\draftnaturalentry{#1}{#2}{#3}{\thepage}}}
\newcommand{\goodoutlinesimple}[2]{\protected@write\outlinefile{}{\string\draftsimpleentry{#1}{#2}{\thepage}}}
\newcommand{\goodoutlinepoint}[3]{\protected@write\outlinefile{}{\string\draftpointentry{#1}{#2}{#3}{\thepage}}}

\makeatother
\newcommand{\goodsetsource}[4]{\def\myfilename{#1}%
   \outlinenatural{0em}{FILE  \underline{{#1}.tex}}{{\bf (#2)} #3 \string\par {\bf STATUS: #4}}
   \pagestyle{myheadings}
   \markright{{#1}.tex #2 \today}}
\newcommand{\goodshowdraft}{%
   \renewcommand{\outlinenatural}[3]{}
   \renewcommand{\outlinepoint}[3]{}
   \renewcommand{\outlinesimple}[2]{}
   \renewcommand{\setsource}[4]{}
   \newpage
   \section{Outline}
   \pagestyle{myheadings}
   \markright{Internally Generated Material, \today}
   \immediate\closeout\outlinefile%
   \par\noindent%
   \begin{list}{}{\setlength{\itemsep}{0pt}\setlength{\labelwidth}{0pt}\setlength{\leftmargin}{0pt}}
      \input{\jobname.outline}
   \end{list}
   \newpage
}

\newlength{\tempwidth}
\newlength{\entrywidth}
\newlength{\symbolwidth}
\newcommand{\draftnaturalentry}[4]{%
   \setlength{\entrywidth}{6in}\addtolength{\entrywidth}{- #1}
   \item \hspace{#1} \parbox[t]{\entrywidth}{\raggedright {\bf #2} \\ #3} \dotfill #4}
\newcommand{\draftpointentry}[4]{%
   \setlength{\entrywidth}{6in}\addtolength{\entrywidth}{- #1}
   \settowidth{\tempwidth}{#2 \hspace{3em}}\addtolength{\entrywidth}{- \tempwidth}
   \item \hspace{#1} {\bf #2} \parbox[t]{\entrywidth}{\raggedright #3} \dotfill #4}
\newcommand{\draftsimpleentry}[3]{%
   \setlength{\entrywidth}{6in}\addtolength{\entrywidth}{- #1}
   \item \hspace{#1} \parbox{\entrywidth}{\raggedright \bf #2} \dotfill #3}

\makeatletter
\newcommand{\makedraft}{%
   \protected@edef\@title{{\tt \Large Draft Material:  Not for Casual Dissemination} \\ \@title}
   \newwrite\outlinefile
   \immediate\openout\outlinefile\jobname.outline
   \typeout{Writing outline file \jobname.outline }%
   \renewcommand{\outlinenatural}{\goodoutlinenatural}
   \renewcommand{\outlinesimple}{\goodoutlinesimple}
   \renewcommand{\outlinepoint}{\goodoutlinepoint}
   \renewcommand{\setsource}{\goodsetsource}
   \renewcommand{\showdraft}{\goodshowdraft}
}
\makeatother

\newtheorem{Theorem}              {Theorem}[section]

\newtheorem{Conjecture}  [Theorem]{Conjecture}

\makeatletter
\newcommand{\numberequationsastheorems}{%
   \renewcommand{\theequation}{\theTheorem}
   \let \c@equation = \c@Theorem}
\makeatother

\newcommand{\formalsynopsis}[2]{%
\outlinepoint{6em}{\string\ref{#1}}{#2 \string\par {Internal Reference:  {\em #1}}}}

\newcommand{\theoremname}{}
\newcommand{\theoremlabel}{}
\newcommand{\theoremdesc}{}
\makeatletter
\renewcommand{\@begintheorem}[5]{%
   \renewcommand{\theoremname}{#1\ #2}
   \renewcommand{\theoremlabel}{\myfilename:#1:#4}
   \renewcommand{\theoremdesc}{#5}
   \label{\theoremlabel:unnamed}%
   \protected@edef\@currentlabel{#1\string~#2}
   \label{\theoremlabel}%
   \begin{samepage}%
   \begin{trivlist}%
      \item[\hskip \labelsep{\bfseries \theoremname}]%
      \itshape}
\renewcommand{\@opargbegintheorem}[6]{%
   \renewcommand{\theoremname}{#1\ #2 [#3]}
   \renewcommand{\theoremlabel}{\myfilename:#1:#5}
   \renewcommand{\theoremdesc}{#6}
   \label{\theoremlabel:unnamed}%
   \protected@edef\@currentlabel{#1\string~#2}
   \label{\theoremlabel}%
   \begin{samepage}%
   \begin{trivlist}%
      \item[\hskip \labelsep{\bfseries \theoremname}]%
      \itshape}
\renewcommand{\@endtheorem}{%
   \end{trivlist} \end{samepage}
   \formalsynopsis{\theoremlabel}{\theoremdesc}}
\makeatother

\makeatletter
\newcommand{\startmyfigurec}[2]{%
   \begin{figure}[thb]%
   \caption{#1}%
   \label{\myfilename:Figure:#2:unnamed}%
   \protected@edef\@currentlabel{Figure\string~\@currentlabel{ [#1]}}
   \label{\myfilename:Figure:#2}%
   \formalsynopsis{\myfilename:Figure:#2}{#1}%
}
\newcommand{\startmyfigure}{\begin{figure}[thb]}
\newcommand{\finishmyfigure}{\end{figure}}
\newcommand{\finishmyfigurebc}[2]{%
   \caption{#1}%
   \label{\myfilename:Figure:#2:unnamed}%
   \protected@edef\@currentlabel{Figure\string~\@currentlabel{ [#1]}}
   \label{\myfilename:Figure:#2}%
   \formalsynopsis{\myfilename:Figure:#2}{#1}%
   \end{figure}}
\makeatother

\newcounter{APointA}[section]
\newcounter{APointB}[APointA]
\newcounter{APointC}[APointB]
\newcounter{APointD}[APointC]
\renewcommand{\theAPointA}{\arabic{APointA}.}
\renewcommand{\theAPointB}{\alph{APointB}.}
\renewcommand{\theAPointC}{\roman{APointC}.}

\makeatletter
\newcommand{\unitsynopsis}[6]{
   \protected@edef\@currentlabel{#3\string~\@currentlabel}
   \label{#5}
   \outlinepoint{#1}{\@currentlabel:}{#3 #4 \string\par {Internal Reference: {\em #5} \string\par {\bf STATUS: #6}}}}
\makeatother

\newcommand{\mysection}[3][unspecified]{\section{#2}%
   \setcounter{APointA}{0}
   \unitsynopsis{1em}{Section \thesection:}{Section}{#2}{\myfilename:Section:#3}{#1}}

\newcommand{\mypoint}[1]{\stepcounter{APointA}%
   \outlinepoint{2em}{\theAPointA}{#1}}
\newcommand{\mysubpoint}[1]{\stepcounter{APointB}%
   \outlinepoint{3em}{\theAPointB}{#1}}
\newcommand{\mysubsubpoint}[1]{\stepcounter{APointC}%
   \outlinepoint{4em}{\theAPointC}{#1}}

\newcounter{AStepA}
\newcounter{AStepB}[AStepA]
\newcounter{AStepC}[AStepB]
\renewcommand{\theAStepB}{\theAStepA.\arabic{AStepB}}
\renewcommand{\theAStepC}{\theAStepB.\arabic{AStepC}}

   {\begin{list}%
   {{\bf Step~\theAStepA}:\hspace{2ex}}%
   {  \usecounter{AStepA}%
      \settowidth{\labelwidth}{{\bf Step~\theAStepA}:\hspace{2ex}}%
      \setlength{\leftmargin}{\labelwidth}%
   }}{\end{list}}
   {\setlength{\leftmargin}{4em} %
   \par\noindent\begin{list}%
   {\hspace{1ex}{\bf Step~\theAStepB}:\hspace{1ex}}%
   {  \usecounter{AStepB}%
      \settowidth{\labelwidth}{{\bf Step~\theAStepB}:\hspace{2ex}}%
      \setlength{\leftmargin}{\labelwidth}%
   }}{\end{list}}
   {\setlength{\leftmargin}{4em} %
   \par\noindent\begin{list}%
   {\hspace{1ex}{\bf Step~\theAStepC}:\hspace{1ex}}%
   {  \usecounter{AStepC}%
      \settowidth{\labelwidth}{{\bf Step~\theAStepC}:\hspace{2ex}}%
      \setlength{\leftmargin}{\labelwidth}%
   }}{\end{list}}
   {\setlength{\leftmargin}{4em} %
   \par\begin{list}%
   {\hspace{1ex}#1\hspace{1ex}}%
   {  
      \settowidth{\labelwidth}{ #1\hspace{2ex}}%
      \setlength{\leftmargin}{\labelwidth}%
   }\item}{\end{list}}



\newenvironment{proof}{%
   \par\noindent{\bf Proof}:~~%
}{%
   \rule{1ex}{1ex}\vspace{1ex}\par%
}

\newcommand{\TBra}     [1]{\{           #1 \}}
\newcommand{\DBkt}     [1]{\left[       #1 \right]}

\renewcommand{\epsilon}{\ensuremath{{\mathbf{\varepsilon}}}}

\newcommand{\Reals}{\ensuremath{{\bf \cal R}}}




%% file: qqbase5.bbl
\begin{thebibliography}{1}

\bibitem[CR]{CR}
J.~Conway and C.~Radin.
\newblock  Quaquaversal tilings and rotations,
\newblock  {\em Inventiones Math.} {\bf 132} (1998), 179-188.


\bibitem[LPS1]{LPS1} A. Lubotsky, R. Phillips and P. Sarnak.
\newblock Hecke Operators and Distributing Points on the Sphere I.
\newblock {\em Comm. Pure Appl. Math.} {\bf 39} (1986) S149-S186.

\bibitem[LPS2]{LPS2} A. Lubotsky, R. Phillips and P. Sarnak.
\newblock Hecke Operators and Distributing Points on $S^2$ II.
\newblock {\em Comm. Pure Appl. Math.} {\bf 40} (1987) 401-420.

\bibitem[R1]{R1}
C.~Radin.
\newblock Symmetry and Tilings, 
\newblock {\em Notices Amer. Math. Soc.} {\bf 42} (1995), 26--31.

\bibitem[R2]{R2} C.~Radin,  private communication.

\bibitem[RS1]{RS1}
C.~Radin and L.~Sadun.
\newblock  Subgroups of
$SO(3)$ associated wtih tilings, 
\newblock {\it J.\ Algebra} {\bf 202} (1998), 611--633

\bibitem[RS2]{RS2}
C.~Radin and L.~Sadun.
\newblock On 2-generator subgroups of $SO(3)$,
\newblock {\it Trans. Amer. Math. Soc}, to appear.

\bibitem[S]{Su} F.~Su, 
\newblock Convergence of random walks on the circle
generated by an irrational rotation.
\newblock {\em Trans. Amer. Math. Soc.} {\bf 350} (1998), 3717-3741.

\end{thebibliography}
